\documentclass[journal]{IEEEtran}
\usepackage{graphicx}
\usepackage{subfigure}
\usepackage{algorithm}
\usepackage{algorithmic}
\usepackage{cite}
\usepackage{amsfonts}
\usepackage{psfrag}
\usepackage{footnote}
\usepackage{amsmath}
\usepackage[switch,pagewise]{lineno}
\usepackage{stfloats}

\newtheorem{lemma}{\textbf{Lemma}}
\newtheorem{theorem}{\textbf{Theorem}}

\begin{document}
%
%
\title{Birkhoff-von-Neumann Switches with Deflection-Compensated Mechanism}

\author{Jinghui Zhang,
        Tong Ye,~\IEEEmembership{Member,~IEEE,}
        Tony T. Lee,~\IEEEmembership{Fellow,~IEEE,}
        Fangfang Yan,
        and Weisheng Hu,~\IEEEmembership{Member,~IEEE}
\thanks{This work was supported by the National Science Foundation of China (61001074, 61172065, and 60825103),
Qualcomm corporation foundation, 973 program (2010CB328205, 2010CB328204), and  Shanghai 09XD1402200.}
\thanks{The authors are with the State Key Laboratory of Advanced Optical Communication
Systems and Networks, Shanghai Jiao Tong University, Shanghai
200030, China. (e-mail: \{zjhrzbb, yetong, ttlee, yff, wshu\}@sjtu.edu.cn)}
}


\maketitle

\begin{abstract}
Despite the high throughput and low complexity achieved by input scheduling based on Birkhoff-von-Neumann (BvN) decomposition; the performance of the BvN switch becomes less predictable when the input traffic is bursty. In this paper, we propose a deflection-compensated BvN (D-BvN) switch architecture to enhance the quasi-static scheduling based on BvN decomposition. The D-BvN switches provide capacity guarantee for virtual circuits (VCs) and deflect bursty traffic when overflow occurs. The deflection scheme is devised to offset the excessive buffer requirement of each VC when input traffic is bursty. The design of our conditional deflection mechanism is based on the fact that it is unlikely that the traffic input to VCs is all bursty at the same time; most likely some starving VCs have spare capacities when some other VCs are in the overflow state. The proposed algorithm makes full use of the spare capacities of those starving VCs to deflect the overflow traffic to other inputs and provide bandwidth for the deflected traffic to re-access the desired VC. Our analysis and simulation show that this deflection-compensated mechanism can support BvN switches to achieve close to 100\% throughput of offered load even with bursty input traffic, and reduces the average end-to-end delay and delay jitter. Also, our result indicates that the packet out-of-sequence probability due to deflection of overflow traffic is negligible, thus only a small re-sequencing buffer is needed at each output port.
\end{abstract}


\IEEEpeerreviewmaketitle

\section{Introduction}\label{intro}
\IEEEPARstart{I}{t} is well known that the throughput of input-queued switches is limited by head-of-line (HOL) blocking. At each time-slot, a traffic scheduler is necessary to configure a connection pattern for the switch fabric to avoid output contentions and maximize the throughput. A good scheduler can improve the system throughput and reduce the end-to-end delay, and should have a low computation complexity. Recently, a number of scheduling algorithms \cite{CSChang:CompComm1,CSChang:CompComm2,TTLEE:JSAC1997,NMcKeown:TON,CSChang:TON2009,Keslassy:INFOCOM2002,BLin:TOC2009,JRLiao:ISITA2010,DLin:ICC2011,CHe:ICC2011,YLi:INFOCOM2001,CSChang:TOC2008,HILee:CL2006,JJJaramillo:ISS2006,XWang:INFOCOM2008,BHu:TON2010,HILee:ICC2008,BHu:JNCA2012,Yshen:HPSR2005,Yshen:HPSR2006,IKeslassy:PhDDis,BLin:TON2010,CSChang:TON2006,BLin:HPI2005}  have been proposed for this purpose.

A class of on-line algorithms \cite{NMcKeown:TON,JRLiao:ISITA2010,DLin:ICC2011,CHe:ICC2011,YLi:INFOCOM2001}, such as iSLIP and dual round-robin matching (DRRM), were devised to compute the input/output paths and rearrange the connection pattern of the switching fabric on the fly according to the requests of input packets. Despite that these on-line algorithms are highly efficient and can achieve 100\% throughput for uniform input traffic, they are difficult to scale when the number of ports $N$ and the line rate are both very large, because these real-time contention resolution algorithms require computation on a slot-by-slot basis, and typically have a complexity on the order of $O(N\log N)$ \cite{Keslassy:INFOCOM2002}.

To avoid real-time computation while providing the capacity guarantee for each input/output (I/O) pair, or virtual circuit (VC), a quasi-static scheduling algorithm, called path switching, was proposed in \cite{TTLEE:JSAC1997}, which was later called Birkhoff-von Neumann (BvN) switching in \cite{CSChang:IWQOS99}. This scheduling algorithm guarantees the capacity assigned for each VC by the repeated executions of a set of predetermined connection patterns, which are calculated from the average loading of all VCs subject to the fixed total switching capacity. The scalability of input-queued switches can be significantly enhanced because this scheduling algorithm does not compute the connection patterns on the fly. Furthermore, results in \cite{BLin:TOC2009} show that the operation overheads, including the computation complexity and memory requirement, can be substantially reduced by the careful design of the switching structure. It was shown in \cite{MCChan:ICC2000} that if the input traffic of each VC is deterministically bounded by a so-called arrival curve \cite{RLCruz:TOIT1991}, this quasi-static algorithm can also achieve 100\% throughput and bounded end-to-end delay. However, the performance of the system becomes less predictable when the input traffic is bursty, mainly due to the explosion of multimedia applications. It is reported in \cite{VWSChan:BCNS2006} that the input traffic at a core router can fluctuate drastically around its mean within a relatively short period of time, though the aggregated traffic is statistically stable in a longer term. In this paper, we propose a compensation mechanism for BvN switches based on the deflection of overflow traffic to cope with the burstiness of arrivals.

\subsection{Previous work on load-balancing and packet out-of -sequence problems in BvN switches}\label{pre-work}

A two-stage load-balanced BvN (LB-BvN) crossbar switch was proposed in \cite{CSChang:CompComm1,CSChang:CompComm2} to deal with such unpredictable traffic fluctuations. In the first stage, the traffic is evenly distributed over all input ports of the second stage, while a set of $N$ circular-shift permutations is periodically running in the second stage to switch the packets to their destinations. Thus, it is essentially a round-robin algorithm that can cope with any traffic conditions \cite{Keslassy:INFOCOM2002,ZZhang:BLOBECOM2011}, and only has a complexity of order $O(1)$. However, immoderate traffic distribution in the first stage introduces a severe packet out-of-sequence problem at the output ports in the LB-BvN switch, even when the input traffic is quite smooth.

One way to fix the out-of-sequence problem is to reorder packets in a so-called re-sequencing buffer at each output port. A Byte-Focal switch is proposed in \cite{Yshen:HPSR2005} to solve the out-of-sequence problem by using the path of each packet along the switch. Though the computation complexity is low, this switch design requires a set of $N^2$ resequencing queues, called virtual input queues in \cite{Yshen:HPSR2005}, at each output port, and the end-to-end delay is on the order of $O(N^2)$ \cite{Yshen:HPSR2006}. The scheme described in \cite{XWang:INFOCOM2008} tried to cope with the out-of-sequence problem by using a three-stage load-balancing switch, which consists of an LB-BvN switch followed by a set of $N$ memory banks as well as an additional switch. This design requires additional hardware and global information exchange for memory reservation in practical implementation \cite{BHu:TON2010}.

Yet another approach to tackle this problem is to prevent packet mis-sequencing in advance. The first effort was made in \cite{CSChang:CompComm2}, where two policies, first-come-first-served (FCFS) and earliest deadline first (EDF), were introduced to manage the central buffers to bound packet out-of-sequence at output buffers. However, the FCFS policy requires the central buffers to speedup $N$ times, while the EDF policy needs to check the timestamps of all packets at each time slot, which makes the system unscalable \cite{CSChang:TOC2008,JJJaramillo:ISS2006}. To reduce the complexity, a class of ¡°frame-based¡± algorithms is proposed in \cite{CSChang:TON2009,Keslassy:INFOCOM2002,JJJaramillo:ISS2006,IKeslassy:PhDDis}, which collect a set of packets of the same flow in either the input buffer or the central buffer so that these packets can be scheduled consecutively as a frame. Obviously, the switches that adopted these frame-based algorithms are not work-conserving \cite{Keslassy:INFOCOM2002}. Furthermore, many existing frame-based algorithms may suffer from various kinds of weaknesses. The Full-Frame First algorithm described in \cite{Keslassy:INFOCOM2002} has a poor scalability because it requires $N^3$ central buffers and global information exchange. The Uniform Frame Spreading algorithm proposed in \cite{IKeslassy:PhDDis} has a poor delay performance even with light traffic as it takes time to accumulate the packets to form a full-frame. The Padded Frame algorithm introduced in \cite{JJJaramillo:ISS2006} underutilizes the bandwidth since it pads dummy packets into the frames to fasten forwarding these frames. Similarly, due to the delivery of fake packets from input buffers to central buffers, the Conservation-and-Reservation algorithm described in \cite{CSChang:TON2009} also cannot fully utilize the bandwidth.

Another kind of scheme is the mailbox switch proposed in \cite{CSChang:TOC2008}, which guarantees in-order packet delivery by tracking the time of each packet departure from the central buffer. However, potential packet contention at the central buffers limits throughput. As mentioned in \cite{CSChang:TON2009}, the throughput of the mailbox switch is low. An improved version of the mailbox switch was the feedback-based LB-BvN switch reported in \cite{HILee:CL2006,BHu:TON2010,BHu:JNCA2012}. Similar to mailbox switches, feedback-based switches require timely feedback of the occupancy status of the central buffers, which imposes a stringent requirement on system implementation. To achieve 100\% throughput in the multi-cabinet case, it is shown in \cite{BHu:TON2010} that such a switch should execute an on-line batch scheduling algorithm with speedup capability at each input port.

In sum, the complexity or cost to reorder out-of-sequencing packets in the LB-BvN switch is quite significant. Actually, packet re-sequencing is equivalent to the function performed by a time-slot interchanger, which possesses the same complexity as a space-division packet switch \cite{TTlee:Book2010}.

\subsection{Our approach in this paper}\label{OurApproach}
In this paper, as mentioned above, we introduce a traffic deflection mechanism to enhance quasi-static scheduling based on BvN decomposition. Deflection can be considered as an alternative buffering scheme, which has been widely used in the packet switching networks. Many packet switch architectures effectively employ deflection routing for resolving contentions among packets in a distributed manner. For example, the tandem-banyan networks \cite{SBassi:TOC1994}, the dual-shuffle exchange networks with error-correcting routing \cite{SCLiew:ICC1992}, and optical burst switches \cite{CFHsu:INFOCOM2002}. Unlike these deflection routing algorithms, the aim of our algorithm is to smooth the fluctuations of input traffic, and to balance the loadings of VCs.

The deflection-compensated BvN (D-BvN) switches provide capacity guarantee for VCs according to the average traffic loading in the same manner as that of BvN switches. However, in contrast to the immoderate traffic distribution in LB-BvN switches, only bursty traffic that overflows from the VCs will be deflected in D-BvN switches. The design of this conditional deflection mechanism is based on the following intrinsic properties of bursty input traffic:

\begin{itemize}
  \item For each individual VC, sometime the burst arrivals may lead to buffer overflow, which implies the same buffer may be empty at some other time and there exists spare capacities not fully utilized.
  \item It is unlikely that the traffic input to all VCs bursts at the same time especially when the number of ports $N$ is large.
\end{itemize}

That is, during a short period of time, some starving VCs have spare capacities while some other VCs are in the overflow state. Our deflection algorithm makes full use of the spare capacities of those starving VCs to cope with the overflow traffic in the following manners:

\begin{itemize}
  \item Some spare capacities act as dynamic buffers, which can be used to deflect the overflow traffic to other inputs such that the overflow traffic can re-access the switching network.
  \item Remaining spare capacities provide bandwidth for the deflected traffic to re-access the desired VC.
\end{itemize}

Our analysis and simulation show that this deflection scheme can support BvN switches to achieve close to 100\% throughput of offered load, and reduce the average end-to-end delay and the delay jitter. Although the deflection also leads to packet out-of-sequence at the outputs, our result indicates that the packet out-of-sequence probability caused by conditional deflections is very small, thus only a small re-sequencing buffer is needed at each output. Thus, this simply designed deflection-compensated mechanism is very easy to implement in practice, yet it has high tolerance to the burst of input traffic.

\begin{figure*}[t]
\centering
\includegraphics[width=0.9\textwidth]{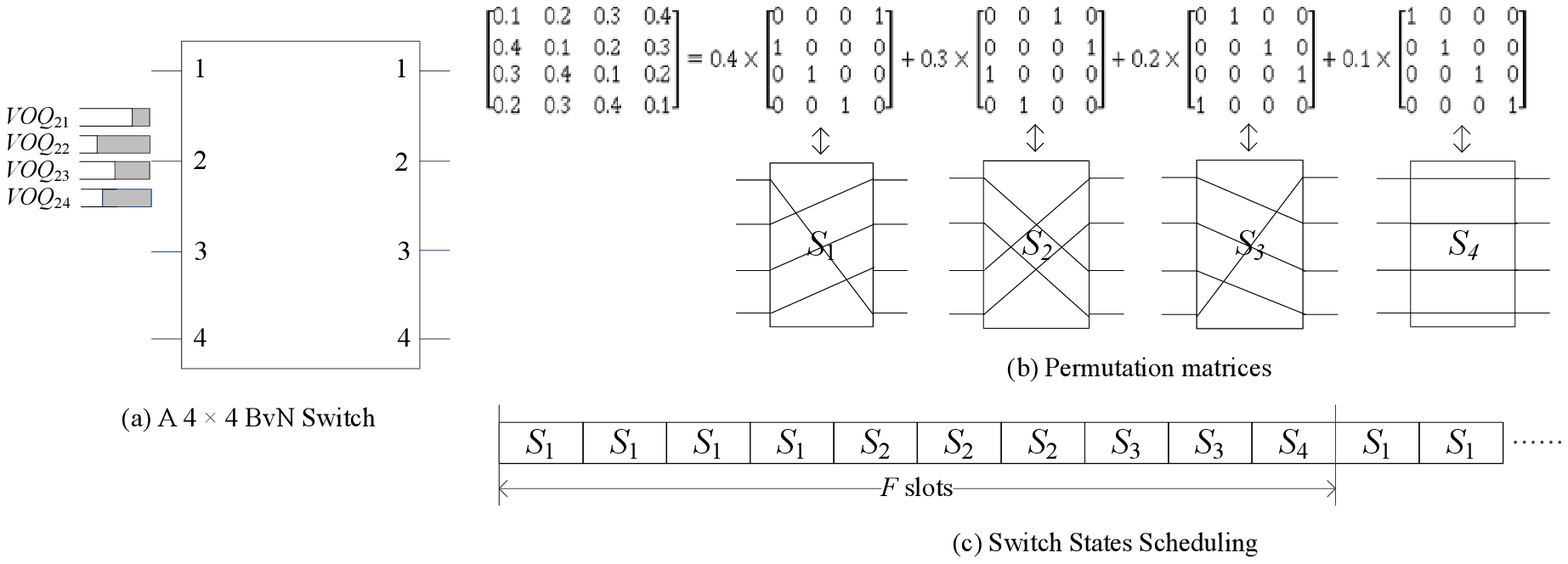}
\caption{Principle of BvN Switches.}\label{BvNswitch}
\end{figure*}

The rest of the paper is organized as follows. In Section \ref{overview}, we briefly introduce the basic concept of BvN switches and then discuss the drawbacks of the quasi-static scheduling. Motivated by the discussion, we propose the D-BvN switch architecture with the deflection-compensated scheduling algorithm. In Section \ref{Model}, the performance of D-BvN switches is analyzed by using the Markov modulated fluid-flow model of input traffic. We devised an ideal deflection approximation technique to estimate the minimum Virtual Output Queue (VOQ) size required to achieve close to 100\% throughput of offered load. In addition, a detailed comparison of the analytical results with simulations is given in this section. To facilitate the presentation, we put the fluid-flow analysis of each VC of D-BvN switches in Appendix \ref{DBvN-ana} and that of BvN switches in Appendix \ref{BvN-ana}, and the simulation model in Appendix \ref{sim-model}. Section \ref{delay-and-jitter} is devoted to the analysis of average end-to-end delay and delay jitter. Section \ref{conclusion} provides the conclusion of this paper.

\section{Preliminaries and Overview}\label{overview}
In this section, we briefly introduce the basic concept of BvN switches, and discuss the relevant limitation of this quasi-static scheduling algorithm, which motivates the deflection-compensated mechanism proposed in this paper.

\begin{figure*}[t]
\centering
\includegraphics[width=0.8\textwidth]{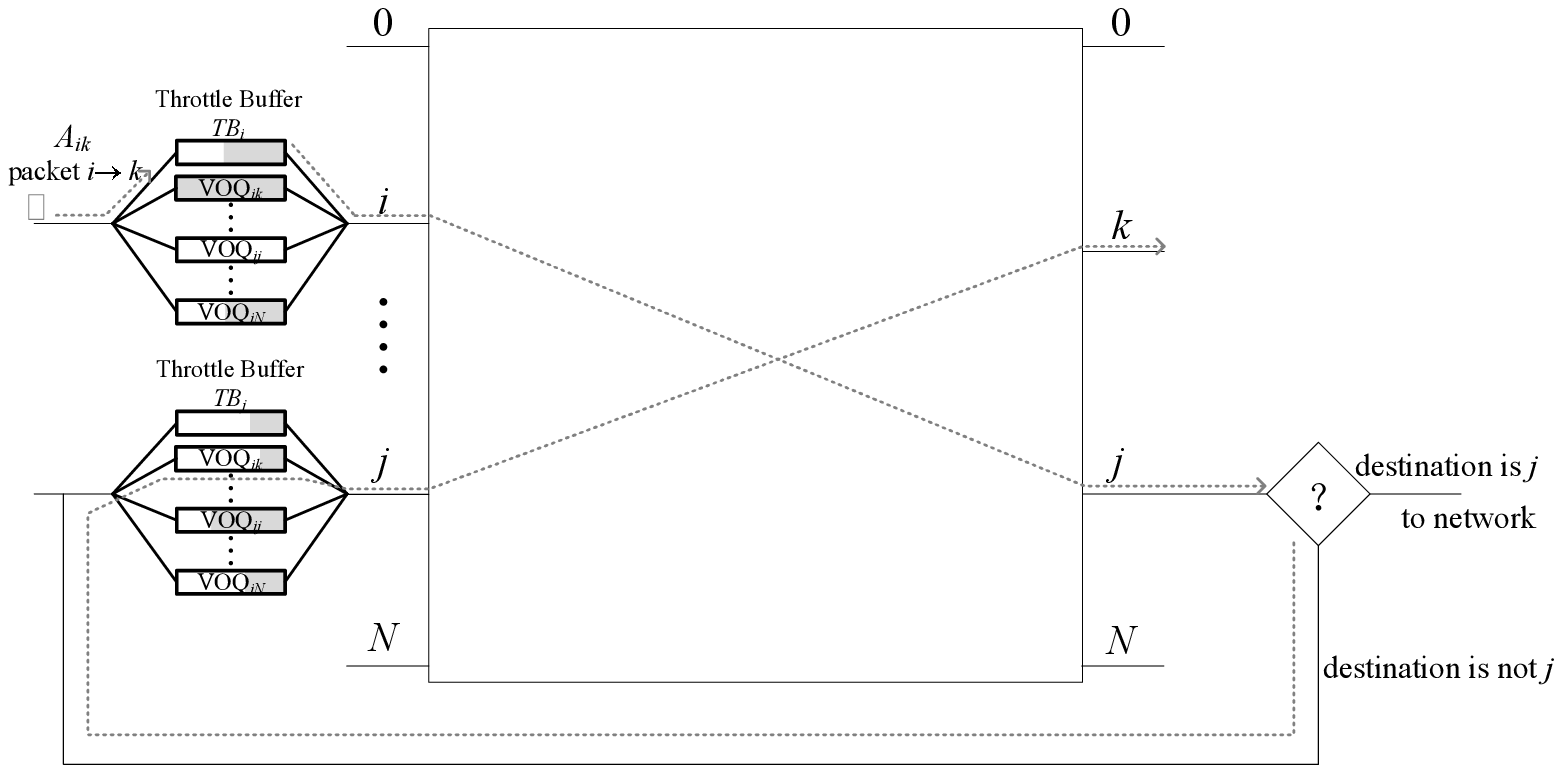}
\caption{Deflection process of an overflow packet.}\label{DBvNswitch}
\end{figure*}

An $N\times N$ BvN switch is plotted in Fig. \ref{BvNswitch}. There are a set of $N$ virtual output queues (VOQs) in each input, in which {VOQ}$_{ij}$ denotes the VOQ at input $i$ to buffer packets destined for output $j$. We assume that the switch fabric under consideration is a crossbar switch to simplify our discussion; the same principle can be extended to a three-stage Clos network in a straightforward manner \cite{TTLEE:JSAC1997}. The quasi-static scheduling algorithm of the BvN switch is based on the Birkhoff-von Neumann decomposition theorem of doubly stochastic matrices \cite{CSChang:INFOCOM2000}.

Given an input traffic matrix $[\lambda_{ij}]_{N\times N}$, where $\lambda_{ij}$ is the average traffic rate from input $i$ to output $j$, and $\sum_{i}\lambda_{ij}\le1$ and $\sum_{j}\lambda_{ij}\le1$, let $[c_{ij}]_{N\times N}$ be a capacity matrix that satisfies the set of constraints $\lambda_{ij}\le c_{ij}$ and $\sum_{i}\lambda_{ij}=\sum_{j}\lambda_{ij}=1$, where $c_{ij}$ is the capacity assigned to the I/O pair $(i,j)$. According to the Birkhoff-von Neumann theorem \cite{CSChang:INFOCOM2000}, the capacity matrix can be decomposed into a sum of permutation matrices and expressed as follows:
\begin{equation}
[c_{ij}]_{N\times N}=\sum_{i=1}^M\phi_i \mathbf{S_i},\label{EQ1}
\end{equation}
where $M$ is an integer less than $N^2-2N+2$, $\mathbf{S_i}$ is a permutation matrix, and the weight $\phi_i>0$ satisfies $\sum_{i=1}^M\phi_i=1$.

Each permutation matrix $\mathbf{S_i}$ represents a connection pattern of the switch, as illustrated in Fig. \ref{BvNswitch}(b). Within a frame of $F$ consecutive time slots, these predetermined connection patterns are scheduled according to their weights in the BvN switch.As the example illustrated in Fig. \ref{BvNswitch}(c), the frame size $F$ is selected to make all $F\phi_i$s integers.In each time slot within a frame, a scheduled permutation establishes $N$ I/O connections for sending packets. For instance, if the permutation $\mathbf{S_1}$ shown in Fig. \ref{BvNswitch}(b) is scheduled in the $k$th time slot in the frame, then input 2 receives a token for sending a packet to output 1 in every $k$th time slot of a frame. According to (\ref{EQ1}), the BvN switch guarantees that the assigned capacity $c_{ij}$ for each I/O pair $(i,j)$, or virtual circuit (VC), can be statistically satisfied by this periodic token assignment scheme.

The Birkhoff-von Neumann decomposition only satisfies the average rate of traffic, and the capacity assignments based on (\ref{EQ1}) are not adaptive to the input traffic fluctuation of each individual VC. Sometimes, the newly arrivals will overflow from the VOQ if the buffer is full due to bursty input traffic. However, the same VOQ could be empty at some other time, and the assigned tokens for the corresponding VC will be discarded. Thus, the main drawback of BvN switches is the throughput degradation in the face of bursty traffic, which not only increases the overflow probability but also underutilizes the assigned capacity.

Since VOQs of these VCs are independent to each other, some VOQs may be full when some other VOQs are empty at the same time, especially if the port number $N$ of the switching fabric is large. This suggests that some spare capacities, called free tokens, of those idle VCs can be utilized to carry the traffic overflow from other VCs to reduce the system loss rate and improve the throughput of the switch. Based on this self-tuning property, we propose a deflection-compensated BvN (D-BvN) scheduling algorithm to relax the limitation of BvN switches.

In a D-BvN switch, the traffic spilled from the saturated VOQs is deflected to the VOQs at other inputs, such that the overflow traffic can re-access the desired VCs. The free tokens play two essential roles in a D-BvN switch: (i) to deflect overflow traffic, and (ii) to provide re-access bandwidth for overflow traffic. At each input port in the D-BvN switch, as plotted in Fig. \ref{DBvNswitch}, a throttle buffer is installed to temporarily store the traffic overflow from saturated VOQs before it is deflected to other inputs. Furthermore, a feedback link is provided between an output and its corresponding input to re-access VCs. As illustrated by the example shown in Fig. \ref{DBvNswitch}, the traffic deflected to output $j$ has another chance to re-access the desired VCs at the input $j$ through the feedback channel between these two ports. This deflection process is illustrated by the example shown in Fig. \ref{DBvNswitch}, in which we adopt the following notations:

\begin{description}[\IEEEsetlabelwidth{VOQ$_{ik}$}\IEEEusemathlabelsep]
  \item[A$_{ik}$] a newly arrived packet at input $i$ that is destined for output $k$
  \item[VC$_{ik}$] the VC between input $i$ and output $k$
  \item[VOQ$_{ik}$] the VOQ that buffers the traffic of VC$_{ik}$
  \item[TB$_i$] the throttle buffer at input $i$
\end{description}

The procedure of deflection is described as follows:

\begin{itemize}[\IEEEsetlabelwidth{Step 3}]
  \item[Step 1] If VOQ$_{ik}$ is not full, A$_{ik}$ enters VOQ$_{ik}$ and waits for the service by VC$_{ik}$; otherwise, A$_{ik}$ joins TB$_i$;
  \item[Step 2] When A$_{ik}$ becomes the HOL packet in TB$_i$ and VC$_{ij}$ currently has a free token (i.e., VOQ$_{ij}$ is empty), A$_{ik}$ is deflected to output $j$ via VC$_{ij}$;
  \item[Step 3] If $j=k$, then A$_{ik}$ reaches its desired output; otherwise, forward A$_{ik}$ back to input $j$ and repeat Step 1.
\end{itemize}
It can be readily seen from the above deflection procedure that the implementation of the D-BvN switch only requires marginally increasing hardware cost and computation complexity.

\section{Fluid-Flow Model of D-BvN Switches}\label{Model}
In this section, the performance of the D-BvN switch is analyzed using a fluid-flow model of input traffic. Our primary goal is to investigate the effectiveness of the deflection-compensated mechanism. Since the deflection by using spare capacity is a scheme designed to offset the VOQ buffer requirements, our analysis initially focuses on the trade-off between the VOQ size and the spare capacity, and aims to estimate the minimum VOQ requirement to achieve the maximum throughput. A widely used approach in the throughput analysis of packet switching systems is to decompose the multi-queue system into independent FIFO queues, and treat each VOQ buffer separately \cite{JYHui:JSAC1987,MJKarol:TOC1987,JYHui:Book1990}. We adopt a similar approach and make the following assumptions in our analysis:

\begin{itemize}
  \item Port number $N$ is large enough such that the deflection scheme can effectively compensate for traffic fluctuations.
  \item To characterize the burstiness of input traffic, we assume that the arrival process of fresh traffic at each VC is a Markov modulated on-off fluid flow \cite{TTlee:Book2010}.
  \item All VCs in the D-BvN switch are statistically identical. That is, all VCs have independent and identical arrival processes, and they possess the same buffer size and equal service capacity.
  \item From the homogenous assumption, the flow conservation law should be held for each VC in a D-BvN switch with large port number $N$. That is, the overflow traffic generated by a VC is equal to that deflected to this VC, and the deflection traffic arriving at each VC is a constant fluid-flow.
\end{itemize}

Suppose the service rate for each input is normalized to 1, and the average arrival rate of fresh traffic is $\bar{\lambda}_p<1$. The parameters and notations used in this paper are listed as follows for easy reference:

\begin{description}[\IEEEsetlabelwidth{$C_2$}]
  \item[$C$] service capacity of each VC, $C=1/N$
  \item[$\bar{\lambda}$] average input traffic rate for each VC, $\bar{\lambda}=\bar{\lambda}_p/N<C$
  \item[$\hat{\lambda}$] peak rate of fresh traffic in the on state, $\hat{\lambda}>C$
  \item[$\alpha$] transition rate of the on state
  \item[$\beta$] transition rate of the off state
  \item[$b$] burstiness of fresh input traffic, $b=1/(\alpha+\beta)$
  \item[$\rho$] offered load of each VC, $\rho=\bar{\lambda}/C$
  \item[$K$] buffer size of VOQ for each VC
  \item[$\Delta$] average rate of the traffic overflow from each VC
  \item[$C_2$] average spare capacity of each VC
\end{description}

Since the BvN decomposition is based on the mean traffic loading, the capacity $C$ assigned to each VC may not be fully utilized because of the burstiness of the input traffic. When the VOQ buffer is empty, the tokens generated according to the assigned capacity $C$ could be wasted because there is no bucket to store them. The spare capacity $C_2$ here is referred to these free tokens that are available to carry packets deflected from other VCs.

In switches based on the BvN scheduling, the maximum throughput cannot exceed the offered load. The throughput of offered load in a D-BvN switch is defined as the ratio of the average rate of the traffic departs from the switch to the average rate of the fresh traffic arriving at the switch. For example, in the fluid-flow model of D-BvN switch under the above assumptions, the maximum throughput of offered load is given by $(C-C_2)/\bar{\lambda}$.

The details of the analysis are described in the rest of this section. In subsection \ref{VOQ-ana}, the behavior of each VC is analyzed by using the fluid-flow model of input traffic \cite{TTlee:Book2010}. We first derive the average rate of overflow traffic spilled from each VC, and then compute the spare capacity that cannot be utilized due to the bursty arrivals. Based on these results, in subsection \ref{Ideal-Def}, an ideal deflection approximation is devised to estimate the minimum VOQ requirement to achieve close to 100\% throughput of offered load. In subsection \ref{Comparison}, we compare the approximate results of loss probability and deflection probability with that obtained by simulations.

\begin{figure}
\centering
\includegraphics[width=0.4\textwidth]{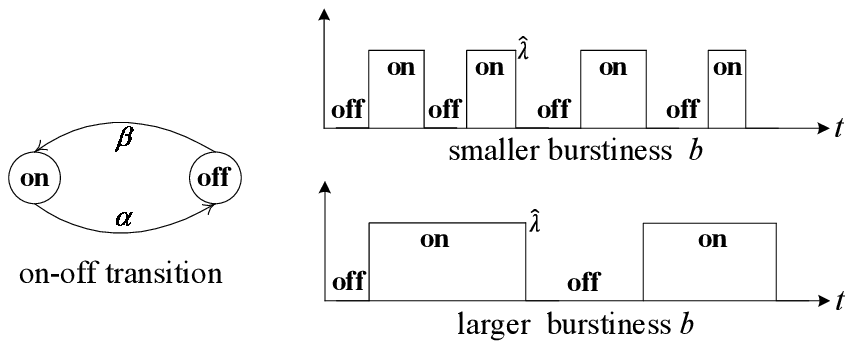}
\caption{The burstiness of on-off fluid flow.}\label{Burstiness}
\end{figure}

\subsection{Analysis of VOQ behavior}\label{VOQ-ana}
We assume that the fresh traffic input to a virtual circuit is a Markov modulated on-off fluid flow, as shown in Fig. \ref{Burstiness}, with an arrival rate $\hat{\lambda}$ in the on state, and $0$ in the off state. The on and off periods are both exponentially distributed with respective mean durations $1/\alpha$ and $1/\beta$. Hence, the peak-to-average ratio of the fresh arrival rate is given by:\[\frac{\hat{\lambda}}{\bar{\lambda}}=\frac{\alpha+\beta}{\beta}.\]
The burstiness $b$ of input traffic is defined as follows:\[b=\frac{1}{\alpha+\beta},\]which is similar to the definition of burst length given in \cite{NMcKeown:TON,RLCruz:TOIT1991}. If the peak-to-average ratio is fixed, then the probability that the input traffic is in an on or off state, denoted as $\pi_{\textrm{on}}=\beta/(\alpha+\beta)$ and $\pi_{\textrm{off}}=1-\pi_{\textrm{on}}$, respectively, is also fixed. It follows that the burstiness $b$ is proportional to the average duration of on state $1/\alpha$, and also the average duration of off state $1/\beta$:\[b=\frac{1}{\alpha}\pi_{\textrm{off}}=\frac{1}{\beta}\pi_{\textrm{on}}.\]
As illustrated in Fig. \ref{Burstiness}, a longer on period in a cycle implies a larger burstiness $b$ in the time domain for the same peak-to-average ratio.

\begin{figure}
\centering
\includegraphics[width=0.45\textwidth]{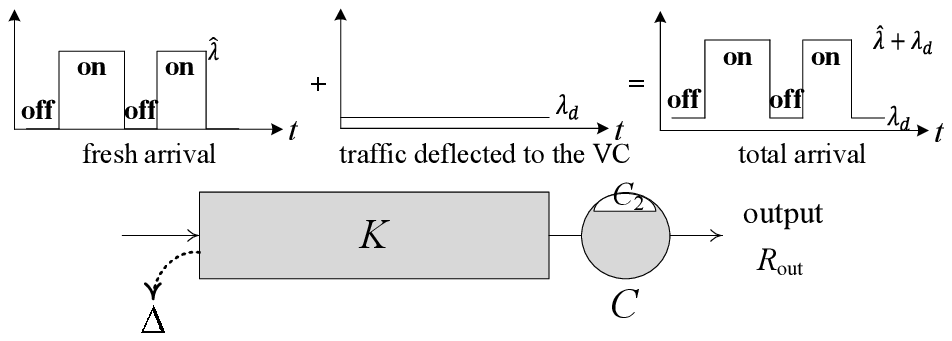}
\caption{The model of a VC with a finite VOQ buffer.}\label{VOQmodel}
\end{figure}

In a D-BvN switch, the superposition of fresh input traffic and overflow traffic deflected from other VCs is fed into each VC. From the assumptions that all VCs are homogeneous and the number of ports $N$ is large, the aggregate traffic deflected to the VC can be considered as a constant fluid flow with intensity $\lambda_d$ in steady state. Therefore, as illustrated in Fig. \ref{VOQmodel}, the superimposed traffic input to each VC is still a Markov modulated on-off fluid flow with a peak rate $\hat{\lambda}+\lambda_d$ in the on state, a deflection rate $\lambda_d$ in the off state, and an average rate $\bar{\lambda}+\lambda_d$.

The overflow packets will be temporarily stored in the throttle buffer, and deflected by the spare capacity, denoted as $C_2$ in Fig. \ref{VOQmodel}. The spare capacity provides the free tokens that are not utilized by idle VCs with empty VOQs. Since the free tokens for traffic deflection are only offered by idle VCs, the capacity $C$ assigned to each VC to serve the input traffic will not be affected by the deflection scheme. As mentioned in Section \ref{overview}, highly burst arrivals not only increase the buffer overflow probability, but also cause poor utilization of assigned capacity $C$. This point is elaborated in the following lemma.

\begin{lemma}
For each VC, the average rate of the overflow traffic is given by
\begin{equation}
\Delta=\textrm{Pr}\{x=K\} \times (\hat{\lambda}+\lambda_d-C),\label{EQ2}
\end{equation}and the average spare capacity is
\begin{equation}
C_2=C-(\bar{\lambda}+\lambda_d)+\Delta,\label{EQ3}
\end{equation}where
\begin{eqnarray}
&&\textrm{Pr}\{x=K\}\nonumber\\
&=&b\times \frac{(\hat{\lambda}+\lambda_d-C)\beta-(C-\lambda_d)\alpha}{(\hat{\lambda}+\lambda_d-C) e^{-\varepsilon K}-(C-\lambda_d)\alpha/\beta} e^{-\varepsilon K}\label{EQ4}
\end{eqnarray}is the probability that the VOQ is full and $\varepsilon$ is a constant defined as follows:
\begin{equation}
\varepsilon=\frac{\alpha}{\hat{\lambda}+\lambda_d-C}-\frac{\beta}{C-\lambda_d}.\label{EQ5}
\end{equation}
\label{lemma1}
\end{lemma}
\begin{IEEEproof}
If the input traffic is in the on state for a long time such that the finite VOQ is saturated, then the new arrivals overflow from the VOQ, as illustrated in Fig. \ref{VOQmodel}. The average rate of the overflow traffic ¦¤ can be calculated by
\begin{eqnarray*}
\Delta&=&\frac{\textrm{overflow traffic in period}\ T}{\textrm{time period}\ T}\\
       &=&\frac{T\times \textrm{Pr}\{x=K\}\times (\hat{\lambda}+\lambda_d-C)}{T}\\
       &=&\textrm{Pr}\{x=K\}\times(\hat{\lambda}+\lambda_d-C).
\end{eqnarray*}
The following probability that the VOQ is full is derived in Appendix \ref{DBvN-ana} based on the standard fluid-flow model described in \cite{TTlee:Book2010}:
\[\textrm{Pr}\{x=K\} =b\times \frac{(\hat{\lambda}+\lambda_d-C)\beta-(C-\lambda_d)\alpha}{(\hat{\lambda}+\lambda_d-C) e^{-\varepsilon K}-(C-\lambda_d)\alpha/\beta} e^{-\varepsilon K},\]
where
\[\varepsilon=\frac{\alpha}{\hat{\lambda}+\lambda_d-C}-\frac{\beta}{C-\lambda_d}.\]
Due to traffic overflow, the average output traffic rate $R_{\textrm{out}}$ is less than the average input traffic rate $\bar{\lambda}+\lambda_d$, and can be calculated by
\begin{eqnarray*}
R_{\textrm{out}}&=&\frac{\textrm{input traffic in period}\ T-\textrm{overflow in period}\ T}{\textrm{time period}\ T} \\
       &=&\frac{(\bar{\lambda}+\lambda_d)T-\Delta T}{T}\\
       &=&\bar{\lambda}+\lambda_d-\Delta.
\end{eqnarray*}
Thus, the average spare capacity that cannot be utilized by the aggregate input traffic is given by
\[C_2=C-R_{\textrm{out}}=C-(\bar{\lambda}+\lambda_d)+\Delta.\]
\end{IEEEproof}

Lemma \ref{lemma1} clearly demonstrates the self-tuning property of the deflection scheme described in Section \ref{overview}. From (\ref{EQ2}) and (\ref{EQ3}), it is easy to see that both the overflow traffic rate $\Delta$ and the spare capacity $C_2$ increase with the burstiness $b$ if the average input traffic rate $\bar{\lambda}+\lambda_d$ and the buffer size $K$ are fixed. This result is consistent with the simulation results shown in Fig. \ref{Selftuning}, where the simulations follow the discrete-time model described in Appendix \ref{sim-model}.

\begin{figure}
\centering
\includegraphics[width=0.45\textwidth]{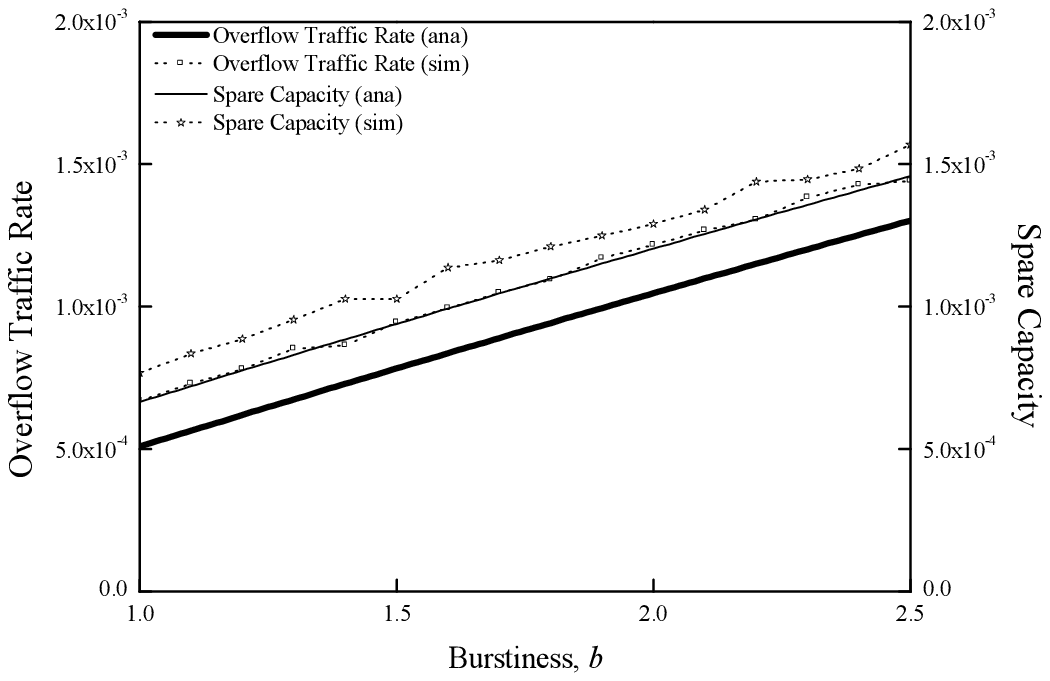}
\caption{Overflow traffic rate and spare capacity versus VOQ size, where $C=1/64$, $\hat{\lambda}=0.8$, $\bar{\lambda}=0.98/64$, $\lambda_d=0.01/64$ and $K=20$.}\label{Selftuning}
\end{figure}

On the other hand, if we fix the burstiness $b$ and decrease the VOQ size $K$, (\ref{EQ2}) and (\ref{EQ3}) of Lemma \ref{lemma1} indicate that both the average rate of the overflow traffic rate $\Delta$ and the average spare capacity $C_2$ will increase simultaneously. In the D-BvN switch, the spare capacity $C_2$ is utilized for the deflection of overflow packets, and serves as a dynamic buffer for the overflow traffic $\Delta$. Thus, there is a trade-off between the real VOQ buffer size $K$, and the ¡°dynamic buffer size¡± $C_2$. In the next subsection, we estimate the appropriate VOQ size by striking a balance between these two buffering strategies.

\subsection{Ideal Deflection Approximation}\label{Ideal-Def}
In this subsection, we focus on the formulation of the trade-off between the VOQ size $K$ and the spare capacity $C_2$. The exact analysis of a feedback stochastic system, such as the D-BvN switch, is mathematically intractable even with the simplified fluid-flow model. Thus, we propose an approximation technique to solve the problem by assuming that the D-BvN switch under consideration is furnished with an ideal deflection mechanism, which guarantees that the spare capacity $NC_2$ is always available to serve overflow packets in the throttle buffer. Apparently, this ideal deflection approximation can only provide some optimistic estimation. Despite some numerical discrepancies, the approximate solutions, in general, agree with simulations, and they can precisely characterize the deflection-compensated scheduling of D-BvN switches. The following measurements are of interest to the throughput analysis of D-BvN switches:
\begin{description}[\IEEEsetlabelwidth{$P_d$}]
  \item[$P_d$] Probability that an input packet is deflected.
  \item[$P_l$] Packet loss probability.
  \item[$\dot{K}$] Minimum VOQ size required to achieve a loss rate of $P_l=10^{-5}$.
\end{description}

In the approximate analysis of the D-BvN switch with an ideal deflection scheme, we use the bold roman type notations $\mathbf{P}_d$ and $\mathbf{P}_l$ to denote the approximations of these probabilities, and $\bf\dot{K}\bf$ to denote the required minimum VOQ size that can achieve 100\% throughput of offered load.

\begin{figure*}[t]
\centering
\includegraphics[width=0.7\textwidth]{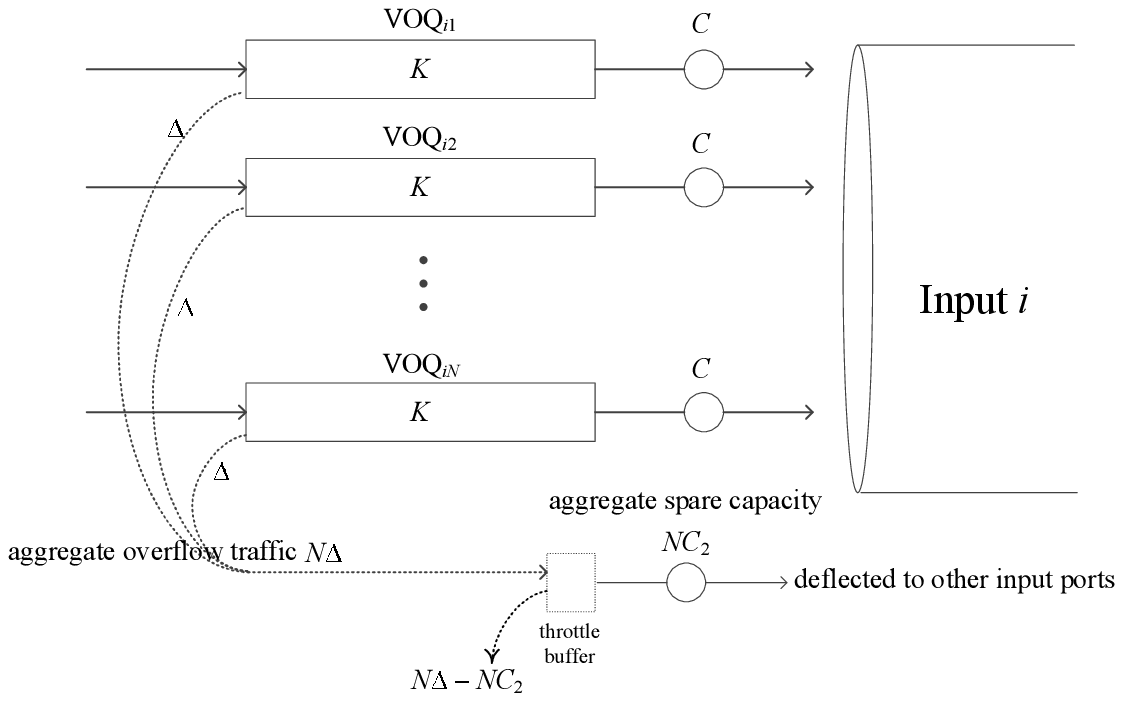}
\caption{The model of an input port with a throttle buffer.}\label{Portmodel}
\end{figure*}

There are $N$ VOQs at each input, as illustrated in Fig. \ref{Portmodel}. Some of these finite VOQs are full while some others are empty at any instant of time due to the burstiness of input traffic. The traffic spilled from these overflow VOQs is fed into the throttle buffer, and then deflected via the spare capacity of idle VCs. For large port number $N$, we consider the overflow traffic input to the throttle buffer is a constant flow $N\Delta$, and the superposition of the spare capacities of all the VCs at an input is also a constant $NC_2$ over time. In this subsection, we show that close to 100\% throughput of offered load can be achieved by a D-BvN switch with a finite VOQ size $\dot{K}$.

We first consider an extreme scenario, where the VOQ size $K$ of each VC is sufficiently small such that the total input rate of a VC is larger than the assigned capacity $C$. From (\ref{EQ3}), we know
\[N(\bar{\lambda}+\lambda_d)>NC\ \textrm{iff}\ N\Delta>NC_2,\]in which case the system is unstable, and the throttle buffer of any finite size will be fully occupied all the time. Under the ideal deflection assumption that all spare capacities $NC_2$ are utilized for deflection, the amount of traffic dropped by the throttle buffer in a period of time $T$ is $(N\Delta-NC_2)T$ at each input. Thus, the traffic loss rate is given by
\begin{equation}
\mathbf{P}_l=\frac{(N\Delta-NC_2)T}{N\bar{\lambda}T}=\frac{\Delta-C_2}{\bar{\lambda}}.\label{EQ6}
\end{equation}
Since the total amount of deflected traffic at each input is $NC_2T$, then each input packet, either a deflected packet or a fresh packet, will be deflected with a probability $\mathbf{P}_d$ given as follows:
\begin{eqnarray}
\mathbf{P}_d&=&\frac{\textrm{traffic deflected at an input in period}\ T}{\textrm{input traffic in period}\ T}\nonumber\\
&=&\frac{NC_2\times T}{N(\bar{\lambda}+\lambda_d)\times T}\nonumber\\
&=&\frac{C_2}{\bar{\lambda}+\lambda_d}.\label{EQ7}
\end{eqnarray}
Since $(\bar{\lambda}+\lambda_d)\mathbf{P}_d$ is the average rate of deflection traffic generated by a VC, it follows from the law of flow conservation, we have
\begin{equation}
\mathbf{P}_d=\frac{\lambda_d}{\bar{\lambda}+\lambda_d}.\label{EQ8}
\end{equation}
From (\ref{EQ7}) and (\ref{EQ8}), we immediately obtain $\lambda_d=C_2$, which is consistent with the assumption that the spare capacity of each VC is fully utilized by the deflection traffic. Furthermore, combining (\ref{EQ2}) and (\ref{EQ3}) under this assumption, we can calculate the average deflection rate $\lambda_d$ from the following equation:
\begin{flalign}
\lambda_d=&C-(\bar{\lambda}+\lambda_d)+(\hat{\lambda}+\lambda_d-C)&\nonumber\\
&\times\frac{\beta}{\alpha+\beta}\frac{(\hat{\lambda}+\lambda_d-C)\beta-(C-\lambda_d)\alpha}{(\hat{\lambda}+\lambda_d-C)\beta e^{-\varepsilon K}-(C-\lambda_d)\alpha}e^{-\varepsilon K}.&\label{EQ9}
\end{flalign}
If $N\Delta>NC_2$, then the packet loss probability $\mathbf{P}_l$ and the deflection probability $\mathbf{P}_d$ can be obtained from (\ref{EQ6}) and (\ref{EQ8}), respectively.

The next scenario, by contrast, we consider a system with a sufficiently large VOQ size $K$, such that the stable condition $N(\bar{\lambda}+\lambda_d)<NC$ always holds, or equivalently, $N\Delta<NC_2$. Again, under the ideal deflection assumption, all overflow traffic can be deflected to other inputs by the spare capacity $NC_2$, and no packet losses, i.e., $\mathbf{P}_l=0$. Therefore, the deflection probability is given by
\begin{equation}
\mathbf{P}_d=\frac{N\Delta \times T}{N(\bar{\lambda}+\lambda_d)\times T}=\frac{\Delta}{\bar{\lambda}+\lambda_d}.\label{EQ10}
\end{equation}
From the law of flow conservation (\ref{EQ8}) and (\ref{EQ10}), we know $\Delta=\lambda_d$. Combining (\ref{EQ2}) and (\ref{EQ3}) again, we can calculate $\lambda_d$ from the following equation
\begin{flalign}
\lambda_d=&(\hat{\lambda}+\lambda_d-C)&\nonumber\\
&\times\frac{\beta}{\alpha+\beta}\frac{(\hat{\lambda}+\lambda_d-C)\beta-(C-\lambda_d)\alpha}{(\hat{\lambda}+\lambda_d-C)\beta e^{-\varepsilon K}-(C-\lambda_d)\alpha}e^{-\varepsilon K}.&\label{EQ11}
\end{flalign}
Then determine the deflection probability $\mathbf{P}_d$ from (\ref{EQ10}) for $N\Delta<NC_2$.

The previous analysis demonstrates that loss of packets occurs in D-BvN switches if the VOQ size $K$ is too small. On the other hand, the spare capacity $NC_2$ will not be fully utilized if the VOQ size $K$ is too large. To determine the minimum VOQ size $K$ without causing packet losses under the ideal deflection assumption, we consider that the system is in the following equilibrium state:
\begin{equation}
N(\bar{\lambda}+\lambda_d)=NC.\label{EQ12}
\end{equation}which is equivalent to $N\Delta=NC_2$. Since the spare capacity $NC_2$ is fully utilized to deflect the overflow traffic $N\Delta$, there is no packet loss. In the following theorem, we show that 100\% throughput of offered load can be achieved by the D-BvN switch with a minimum VOQ size $\mathbf{\dot{K}}$, and the packet deflection probability $\mathbf{\dot{P}}_d$ is independent of the burstiness $b$.

\begin{theorem}\label{theorem1}
Under the ideal deflection assumption, the D-BvN switch with the following VOQ size for each VC can achieve 100\% throughput of offered load:
\begin{equation}
\mathbf{\dot{K}}=b\bar{\lambda}\times \left[\frac{\alpha}{\beta}\left(\frac{\rho}{1-\rho}-1\right)-1\right],\label{EQ13}
\end{equation}and the deflection probability is given by
\begin{equation}
\mathbf{\dot{P}}_d=1-\rho.\label{EQ14}
\end{equation}
\end{theorem}
\begin{IEEEproof}
Under the ideal deflection assumption, if the D-BvN switch is in the equilibrium state $N\Delta=NC_2$, i.e., $\bar{\lambda}+\lambda_d=C$, both (\ref{EQ9}) and (\ref{EQ11}) become the following equation:
\[\frac{\beta(\hat{\lambda}-\bar{\lambda})}{\beta\left\{1+\mathbf{\dot{K}}(\hat{\lambda}-\bar{\lambda})\times \left[\frac{\alpha}{(\hat{\lambda}-\bar{\lambda})^2}+\frac{\beta}{\bar{\lambda}^2}\right]\right\}+\alpha}=C-\bar{\lambda},\]
from which we immediately obtain the expression (\ref{EQ13}) of $\mathbf{\dot{K}}$. Similarly, the deflection probability $\mathbf{\dot{P}}_d$ is obtained by substituting (\ref{EQ12}) into (\ref{EQ8}).
\end{IEEEproof}

In the equilibrium state where $N\Delta=NC_2$, or $N(\bar{\lambda}+\lambda_d)=NC$, and under the ideal deflection assumption, the spare capacity $NC_2$ is fully employed as the ¡°dynamic buffer¡± to accommodate the overflow traffic $N\Delta$. The optimality of VOQ size $\mathbf{\dot{K}}$ given in (\ref{EQ13}) can be interpreted by (\ref{EQ14}), which indicates that the VC is busy all the time: either busy in transmitting a packet with probability $\rho$, or deflecting a packet with probability $\mathbf{\dot{P}}_d=1-\rho$.

Theorem \ref{theorem1} also implies that the D-BvN switch with finite VOQ size $\mathbf{\dot{K}}$ can achieve 100\% throughput of offered load with packet loss rate $\mathbf{P}_l=0$. However, this perfect performance of the D-BvN switch with the ideal deflection mechanism can never be realized in practice. Instead, we define $\dot{K}$ to be the required minimum VOQ size that a D-BvN switch can have to achieve a loss rate of $P_l=10^{-5}$ or, equivalently, a throughput very close to 100\%. In general, this critical VOQ size $\dot{K}$ in practice is larger than $\mathbf{\dot{K}}$, but the gap can be reduced by increasing the throttle buffer size. A detailed comparison of the ideal deflection approximation and the simulation results is provided in the next subsection.

\subsection{Comparison with Simulation Results}\label{Comparison}
In the simulation, the time is slotted and the input traffic of each VC is a discrete on-off Markov modulated process with parameters described in Appendix \ref{sim-model}. In addition, the deflection mechanism is modeled by discrete stochastic events in our simulation. The assumption of an ideal deflection compensation mechanism is only adopted in the approximation, which provides an optimistic lower bound as we mentioned before. For the purpose of performance comparisons, we also provide the analytical and simulation results of the BvN switch. The analytical results are given in Appendix \ref{BvN-ana}.

\subsubsection{Packet loss probability and optimal VOQ size}\label{kc-pl-comp}
\begin{figure}
\centering
\includegraphics[width=0.45\textwidth]{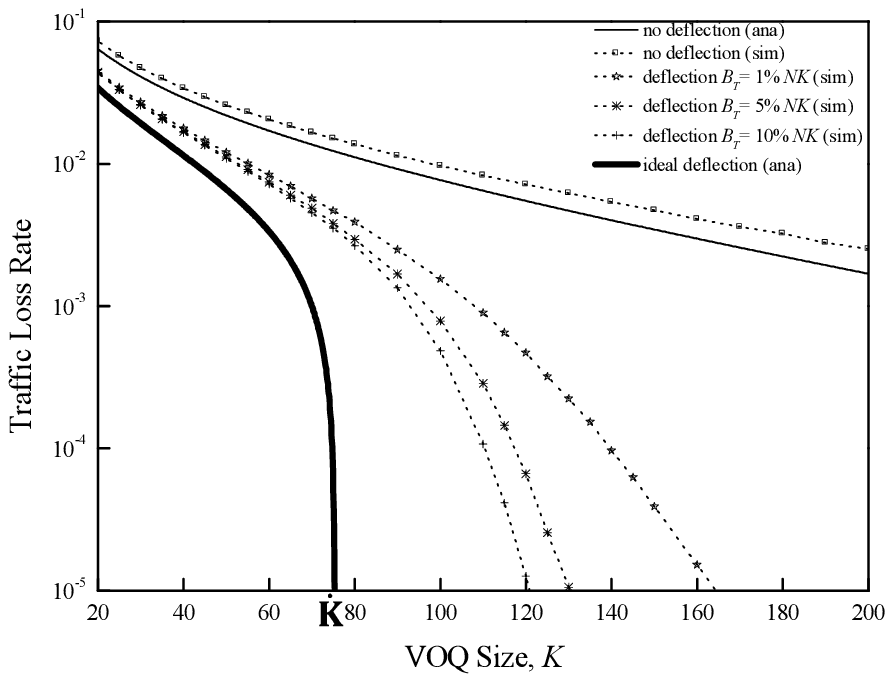}
\caption{Traffic loss rate versus VOQ size, where $N=64$, $\hat{\lambda}=0.8$, $\bar{\lambda}=0.98/64$, $\alpha=0.49$, $\beta=0.0096$ and $b=2$.}\label{LossRate}
\end{figure}

Both analytical and simulation results of the packet loss probability $P_l$ versus VOQ size $K$ are plotted in Fig. \ref{LossRate}, where all simulation results of $P_l$ are lower-bounded by the approximate loss probability $\mathbf{P}_l$ derived under the ideal deflection assumption, and upper-bounded by that of the BvN switch without deflection. Herein, the analytical loss rate of the BvN switch is plotted according to (\ref{EQ29}) given in Appendix \ref{BvN-ana}. The throttle buffer size of each simulation curve displayed in Fig. \ref{LossRate} is specified by ¡°$B_T=x\% NK$¡±, which means the throttle buffer size is equal to $x\%$ of the size of all VOQs at each input.

In Fig. \ref{LossRate}, the loss probability of D-BvN with ideal deflection assumption quickly approaches $0$ when the VOQ size is larger than a critical value $\mathbf{\dot{K}}$, which suggests that 100\% throughput of offered load can be achieved by D-BvN switches with a finite VOQ size $\mathbf{\dot{K}}$. However, all simulation results presented in Fig. \ref{LossRate} also show that the loss probability will never drop to $0$ if the VOQ size $K$ is finite. Nevertheless, the system still can achieve close to 100\% throughput of offered load with a finite VOQ size $\dot{K}$. For example, the critical value $\dot{K}$ shown in Fig. \ref{LossRate} lies in the range between $1.5\mathbf{\dot{K}}$ and $2\mathbf{\dot{K}}$ when $P_l=10^{-5}$. The discrepancy between the simulation and analytical results is due to the following reasons:

\begin{itemize}
  \item In the fluid-flow analysis, when the port number $N$ is large, we assume that the input to the throttle buffer is a constant flow, and the available spare capacity is also constant over time. However, both of them are discrete probabilistic events in the simulation model.
  \item The approximation is obtained under the ideal deflection assumption, such that the utilization of spare capacity for deflection is maximized. But the arrivals of overflow packets and the availability of free tokens are both probabilistic and the size of throttle buffer is finite in the discrete stochastic model of simulation. Thus, the services provided by free tokens to deflect packets may not always ready. However, the gap between $P_l$ and $\mathbf{P}_l$ can be reduced by increasing the throttle buffer size as shown in Fig. \ref{LossRate}.
\end{itemize}

In spite of these numerical discrepancies, our analytical results clearly demonstrate the characteristic of D-BvN switches. In comparison with the BvN switch without deflection, it can be seen from Fig. \ref{LossRate} that the packet loss probability of the BvN switch is significantly higher than that of the D-BvN switch, which decreases slowly with the increasing of VOQ size $K$. In other words, without the deflection compensation mechanism, the BvN switch requires a much larger VOQ buffer to achieve the same throughput of offered load.

\begin{figure}
\centering
\includegraphics[width=0.45\textwidth]{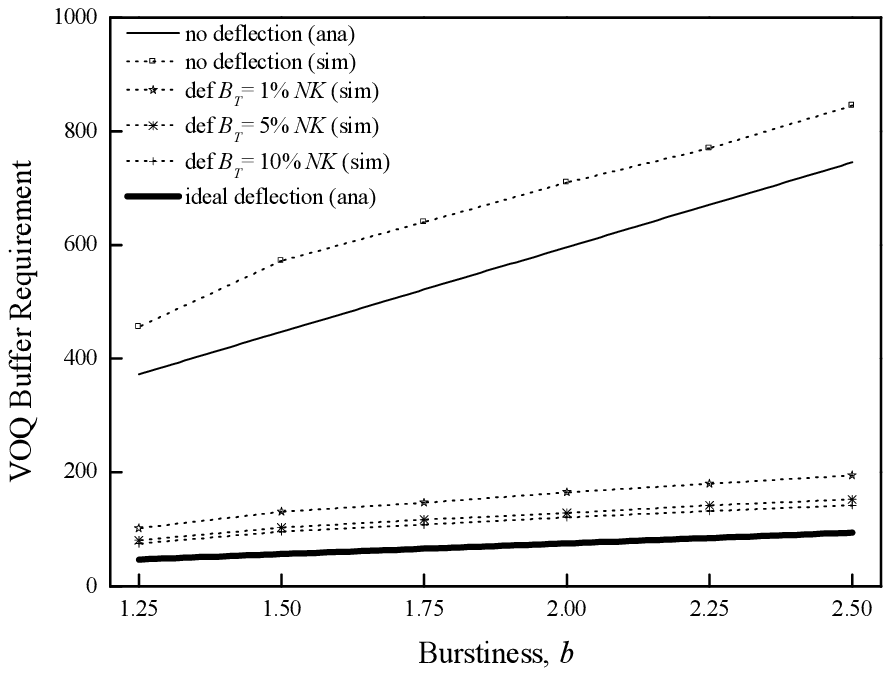}
\caption{Minimum VOQ size versus burstiness, where $N=64$, $\hat{\lambda}=0.8$, and $\bar{\lambda}=0.98/64$.}\label{VOQRequirement}
\end{figure}

Presumably, according to the BvN scheduling, the VOQ buffer is almost unnecessary if the traffic input to each VC is smooth enough and satisfies the stable condition $\bar{\lambda}<C$; the buffer is indispensable only when the input traffic is bursty. This intuitive interpretation of the VOQ function is consistent with (\ref{EQ13}) of the above theorem, in which the required VOQ size $\mathbf{\dot{K}}$ is linearly proportional to the burstiness $b$. As shown in Fig. \ref{VOQRequirement}, the same property holds for the critical value $\dot{K}$ that is obtained by the simulation. Fig. \ref{VOQRequirement} also reveals that a much larger VOQ size is required by the BvN switch without deflection to achieve the same throughput of offered load, in which the analytical curve of the BvN switch is given by (\ref{EQ29}) in Appendix \ref{BvN-ana}.

\subsubsection{Packet deflection probability and out-of-sequence problem}\label{pd-prob}
\begin{figure}
\centering
\includegraphics[width=0.45\textwidth]{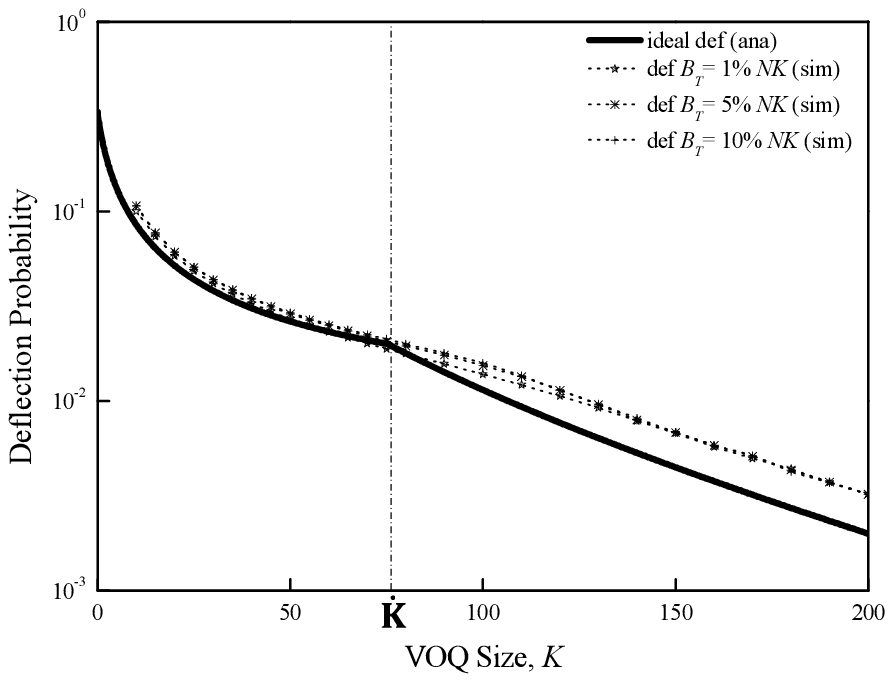}
\caption{Packet deflection probability versus VOQ size, where $N=64$, $\hat{\lambda}=0.8$, $\bar{\lambda}=0.98/64$, $\alpha=0.49$, $\beta=0.0096$ and $b=2$.}\label{PDvsK}
\end{figure}

The deflection probability $P_d$ obtained by simulations and the approximations $\mathbf{P}_d$ given by (\ref{EQ8}) for $K\le\mathbf{\dot{K}}$, and (\ref{EQ10}) for $K>\mathbf{\dot{K}}$ are all displayed in Fig. \ref{PDvsK}. We note that in general, $P_d>\mathbf{P}_d$ for all VOQ size $K$ and any throttle buffer size $B_T$. Of course, the difference is caused by the ideal deflection assumption in the analysis, because additional packet losses may occur in the simulation.

\begin{figure}
\centering
\includegraphics[width=0.45\textwidth]{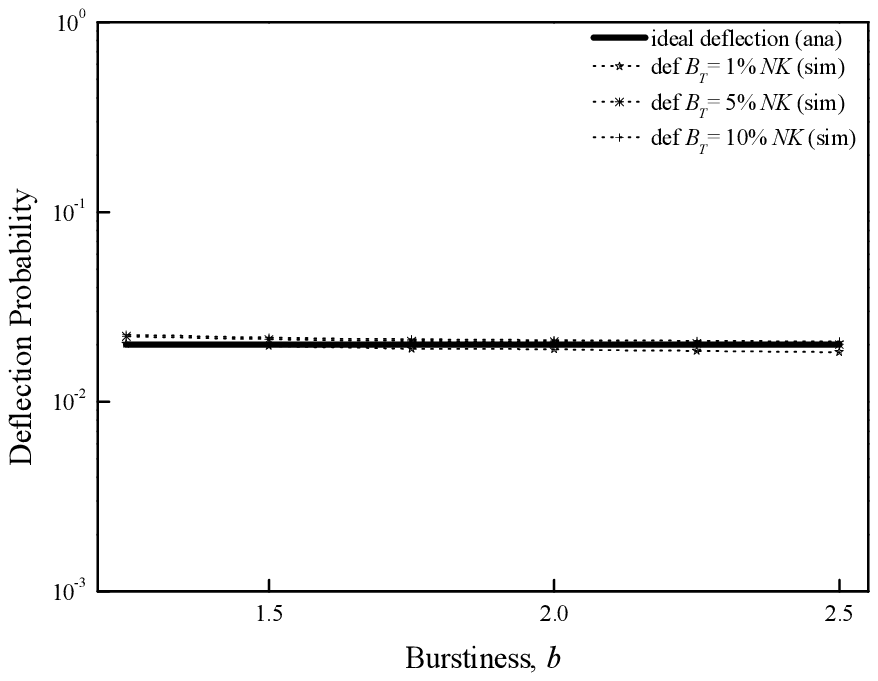}
\caption{Deflection probability versus burstiness, where $N=64$, $\hat{\lambda}=0.8$, $\bar{\lambda}=0.98/64$ and $K=\dot{K}$ or $\mathbf{\dot{K}}$.}\label{PDvsB}
\end{figure}

The packet out-of-sequence problem is unavoidable in the D-BvN switch due to packet deflections. Since the capacity of each VC is guaranteed by BvN scheduling, input packets should normally be routed by the VC within the scheduled time. Thus, a fresh input packet without experiencing any deflections should reach the desired output in the same sequential order of input. That is, out-of-sequence only occurs to deflected packets. Since a fresh packet could be deflected in the D-BvN switch with a probability $P_d$, the packet out-of-sequence probability should be upper-bounded by this deflection probability $P_d$. As shown in Fig. \ref{PDvsK}, the discrepancy between $P_d$ and $\mathbf{P}_d$ is almost negligible, thus the packet out-of-order probability should be in the same order as that of $\mathbf{P}_d$. Furthermore, in the equilibrium state, Theorem \ref{theorem1} indicates that deflection probability $\mathbf{\dot{P}}_d$ is completely determined by the offered load $\rho$ and independent of the burstiness $b$. This point is reinforced by the simulations shown in Fig. \ref{PDvsB}, where, again, the difference between $P_d$ and $\mathbf{P}_d$ is insignificant.

\section{End-to-end Delay and Delay Jitter}\label{delay-and-jitter}
In this section, we first derive the average end-to-end delay and delay jitter of D-BvN switches under the ideal deflection assumption, and then compare these analytical results with simulations. Again, we show that the delay performances of D-BvN switches are upper-bounded by that of BvN switches without deflections, and lower-bounded by that with ideal deflections. The following parameters and notations are used in the delay analysis of D-BvN switches with the ideal deflection mechanism:

\begin{description}[\IEEEsetlabelwidth{$\mathbf{D}_q$}]
  \item[$\mathbf{D}$] end-to-end packet delay
  \item[$\mathbf{D}_q$] queuing delay in the VOQ buffer
  \item[$a$] a constant cross-switch delay
\end{description}

\subsection{Analysis of Delay and Jitter}\label{delay-jitter-ana}
In the delay analysis, we consider that the D-BvN switch under study is operated in the stable region, i.e., $N(\bar{\lambda}+\lambda_d)\le NC$, in which no packet losses, i.e., $\mathbf{P}_l=0$, will occur under the ideal deflection assumption. Each packet input to a VC will either be deflected to another input with probability $\mathbf{P}_d$ if the VOQ is full, or join the VOQ with probability $1-\mathbf{P}_d$. Thus, the average end-to-end delay $E[\mathbf{D}]$ can be expressed as follows:
\begin{flalign}
E[\mathbf{D}]=&E[\textrm{delay if it is deflected}]\times \mathbf{P}_d&\nonumber\\
&+E[\textrm{delay if no deflection}]\times (1-\mathbf{P}_d)&\nonumber\\
=&\{\textrm{cross switch delay}+E[\mathbf{D}]\}\times \mathbf{P}_d&\nonumber\\
&+E[\textrm{Queuing delay in VOQ}]\times (1-\mathbf{P}_d)&\nonumber\\
=&E[a+\mathbf{D}]\mathbf{P_d}+E[\mathbf{D}_q](1-\mathbf{P}_d).&\label{EQ15}
\end{flalign}
It follows that
\begin{equation}
E[\mathbf{D}]=\frac{a\mathbf{P}_d}{1-\mathbf{P}_d}+E[\mathbf{D}_q].\label{EQ16}
\end{equation}where the first term is the average delay incurred by deflections in the D-BvN switch, and the second term, mean queuing delay $E[\mathbf{D}_q]$, is given by (\ref{EQ24}) in Appendix \ref{DBvN-ana}. The mean delay $E[\mathbf{D}]$ can be determined by simultaneously solving (\ref{EQ16}) together with (\ref{EQ10}) and (\ref{EQ11}). Since the deflection probability $\mathbf{P}_d$ is typically very small in the stable region, the mean delay $E[\mathbf{D}]$ in practice is predominated by the queuing delay spent by the packet waiting in the VOQ buffer.

The delay jitter of a packet is defined by the variance $var[\mathbf{D}]$  of the end-to-end delay $\mathbf{D}$. As with the recursive relation (\ref{EQ15}) for the mean delay, the second moment $E[\mathbf{D}^2]$ can be evaluated in a similar manner as follows:
\begin{equation}
E[\mathbf{D}^2]=E[\mathbf{D}_q^2](1-\mathbf{P}_d)+E[(a+\mathbf{D})^2]\mathbf{P}_d.\label{EQ17}
\end{equation}or equivalently, we have
\[E[\mathbf{D}^2]=E[\mathbf{D}_q^2]+\frac{\{a^2+2aE[\mathbf{D}]\}\times \mathbf{P}_d}{1-\mathbf{P}_d}.\]
Thus, the delay jitter is given by
\begin{eqnarray}
var[\mathbf{D}]&=&E[\mathbf{D}^2]-E^2[\mathbf{D}]\nonumber\\
&=&E[\mathbf{D}_q^2]-E^2[\mathbf{D}_q]+a^2\frac{\mathbf{P}_d}{(1-\mathbf{P}_d)^2}\nonumber\\
&=&var[\mathbf{D}_q]+a^2\frac{\mathbf{P}_d}{(1-\mathbf{P}_d)^2}.\label{EQ18}
\end{eqnarray}
where the second moment of queuing delay $E[\mathbf{D}_q^2]$ is given by (\ref{EQ25}) in Appendix \ref{DBvN-ana}. Again, the delay jitter $var[\mathbf{D}]$ can be simultaneously solved by (\ref{EQ10}), (\ref{EQ11}), and (\ref{EQ18}), and it is typically predominated by the variance of queuing delay $var[\mathbf{D}_q]$ in the VOQ buffer.

When the D-BvN switch is in the equilibrium state, $N(\bar{\lambda}+\lambda_d)=NC$, the end-to-end delay and the queuing delay in the VOQ buffer will be denoted by $\mathbf{\dot{D}}$ and $\mathbf{\dot{D}}_q$, respectively. In this particular case, we obtain the following explicit expressions of delay and delay jitter.

\begin{theorem}\label{theorem2}
In the equilibrium state, $N(\bar{\lambda}+\lambda_d)=NC$, the average end-to-end delay and the delay jitter in a D-BvN switch with ideal deflection mechanism are given by
\begin{equation}
E[\mathbf{\dot{D}}]=a\times \frac{1-\rho}{\rho}+E[\mathbf{\dot{D}}_q],\label{EQ19}
\end{equation}
and
\begin{equation}
var[\mathbf{\dot{D}}]=a^2\times \frac{1-\rho}{\rho^2}+var[\mathbf{\dot{D}}_q],\label{EQ20}
\end{equation}
respectively, where
\begin{flalign}
E[\mathbf{\dot{D}}_q]&=b\times&\nonumber\\
&\frac{\left[\left(2\frac{\alpha}{\beta}+1\right)\rho-\left(1+\frac{\alpha}{\beta}\right)\right]\left[\left(2-\frac{\beta}{\alpha}\right)\rho+\left(\frac{\beta}{\alpha}-1\right)\right]}{2\rho(1-\rho)},&\nonumber\\
\label{EQ21}
\end{flalign}
and
\begin{flalign}
E[\mathbf{\dot{D}}_q^2]=&b^2\times&\nonumber\\ &\frac{\left[\left(2\rho-1\right)\frac{\alpha}{\beta}-\left(1-\rho\right)\right]^2\left[\left(1-\rho\right)\frac{2\beta}{\alpha}+2\rho-1\right]}{3(1-\rho)^2}.&\nonumber\\
\label{EQ22}
\end{flalign}
\end{theorem}
\begin{IEEEproof}
When the D-BvN switch is in the equilibrium state, $N(\bar{\lambda}+\lambda_d)=NC$, we know from Theorem \ref{theorem1} that the deflection probability is given by $\mathbf{\dot{P}}_d=1-\rho$. Thus, from (\ref{EQ16}), (\ref{EQ18}), (\ref{EQ24}), and (\ref{EQ25}), we immediately obtain the results (\ref{EQ21}) and (\ref{EQ22}).
\end{IEEEproof}

Theorem \ref{theorem2} demonstrates that a packet input to a D-BvN switch in the equilibrium state could be deflected $(1-\rho)/\rho$ times on the average before it is switched by the VC to its desired output. Thus, both the mean delay and delay jitter incurred by deflections are independent of the burstiness $b$. This characteristic of D-BvN switches in the equilibrium state can be attributed to the fact that the deflection probability $1-\rho$ is independent of $b$, which was elaborated in Theorem \ref{theorem1} in Section \ref{Ideal-Def}.

\subsection{Comparison with Simulation Results}\label{delay-comparison}
In this subsection, we provide a comparison of the analytic results of delay and delay jitter under the ideal deflection assumption with simulation results. Recall that the required critical VOQ size for the D-BvN switch to achieve a loss rate of $P_l=10^{-5}$ is $\dot{K}$. Let $\dot{D}$ be the end-to-end delay of a packet in the D-BvN switch with VOQ size $\dot{K}$.

\begin{figure}
\centering
\includegraphics[width=0.45\textwidth]{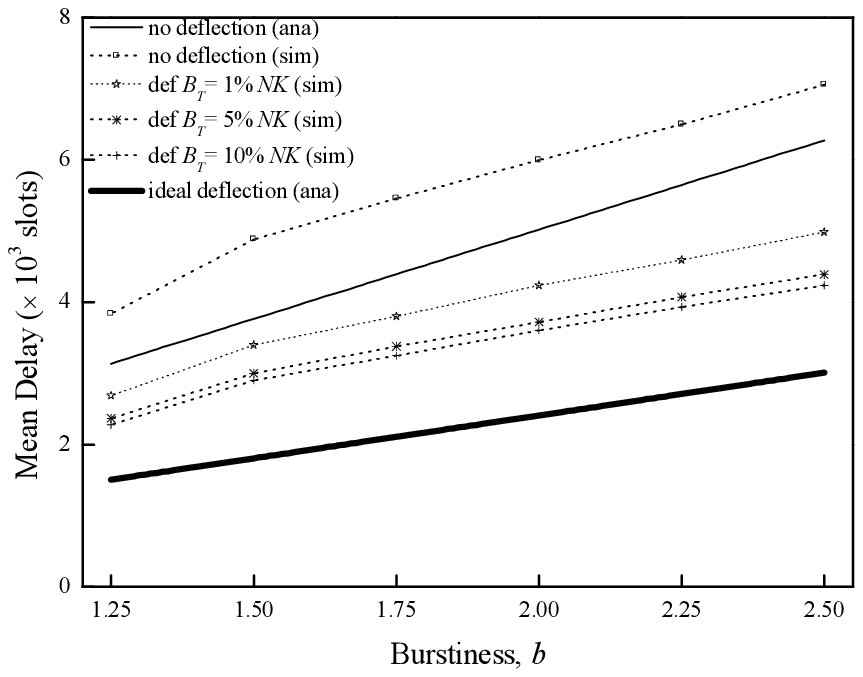}
\caption{Mean delay versus burstiness, where $N=64$, $\hat{\lambda}=0.8$, $\bar{\lambda}=0.98/64$ and $K=\dot{K}$ or $\mathbf{\dot{K}}$.}\label{Delay}
\end{figure}

The average end-to-end delay versus burstiness is plotted in Fig. \ref{Delay}, where the average delay $E[\mathbf{\dot{D}}]$ given by (\ref{EQ16}) is compared with the simulation result $E[\dot{D}]$. Recall that the capacity of each input port is equally divided and assigned to $N$ VCs in our model, which implies that each VC only receives one token in a frame of $N$ time slots. Thus, the average packet queuing delay shown in Fig. \ref{Delay} is on the order of $N$ times the mean queue length of the VOQ, which is consistent with the results of BvN switches reported in \cite{CSChang:CompComm1}. All curves in Fig. \ref{Delay} demonstrate that the average end-to-end delay linearly increases with the burstiness $b$. Furthermore, we also provide the delay performance of the BvN switch when the packet loss rate is $10^{-5}$ in Fig. \ref{Delay}, where the analytical curve is plotted according to (\ref{EQ30}) in Appendix \ref{BvN-ana}. As we mentioned before, Fig. \ref{Delay} shows that all simulation results of $E[\dot{D}]$ are lower-bounded by $E[\mathbf{\dot{D}}]$, which is derived in the equilibrium state under the ideal deflection assumption, and outperforms that of the BvN switch without deflections.

As shown in Fig. \ref{LossRate} and \ref{VOQRequirement}, a very large VOQ buffer is required to achieve a traffic loss rate of $10^{-5}$ in the BvN switch. Thus, we should expect that backlogged packets will form a long queue in the VOQ buffer. However, a much smaller VOQ buffer size $\dot{K}$ is enough for each VC in the D-BvN switch to achieve the same loss rate. Therefore, according to Little's law in queuing theory, the queuing delay in a D-BvN switch should be much less than that in a BvN switch for achieving the same throughput of offered load.

\begin{figure}
\centering
\includegraphics[width=0.45\textwidth]{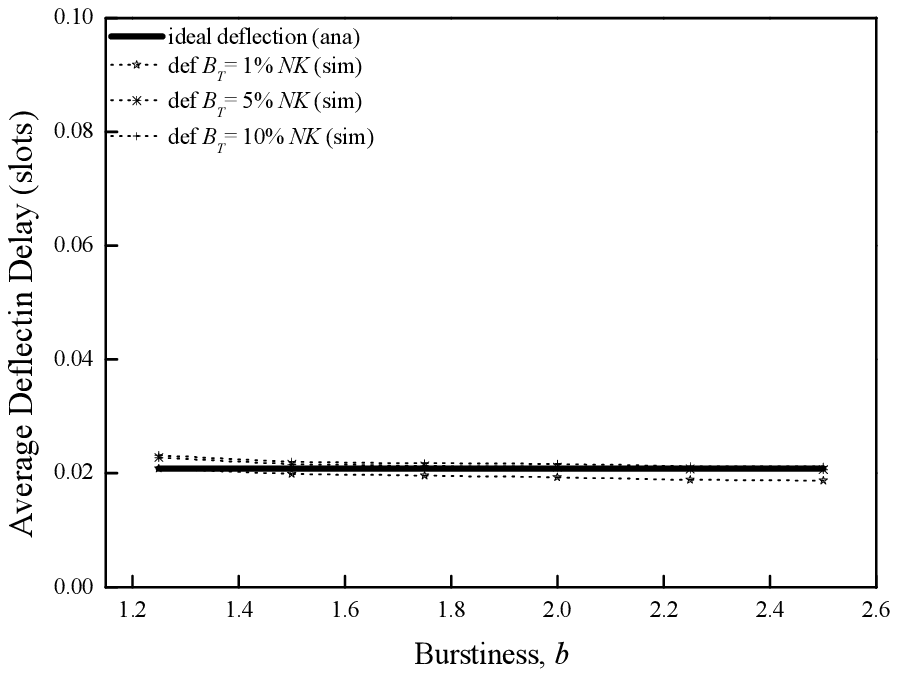}
\caption{Mean deflection delay versus burstiness, where $N=64$, $\hat{\lambda}=0.8$, $\bar{\lambda}=0.98/64$ and $K=\dot{K}$ or $\mathbf{\dot{K}}$.}\label{DefDelay}
\end{figure}

In Fig. \ref{DefDelay}, the average deflection delay in expression (\ref{EQ19}) is compared with the counterpart obtained by the simulation model. All curves confirm that the average deflection delay is independent of the burstiness $b$. In sum, the deflection mechanism of a D-BvN switch allows sharing of spare capacities among different VCs at the expense of negligible deflection delay. This simply designed scheme not only can reduce the VOQ buffer requirement, but also significantly improves the delay performances of the BvN switches, especially in the face of highly bursty input traffic.

\begin{figure}
\centering
\includegraphics[width=0.45\textwidth]{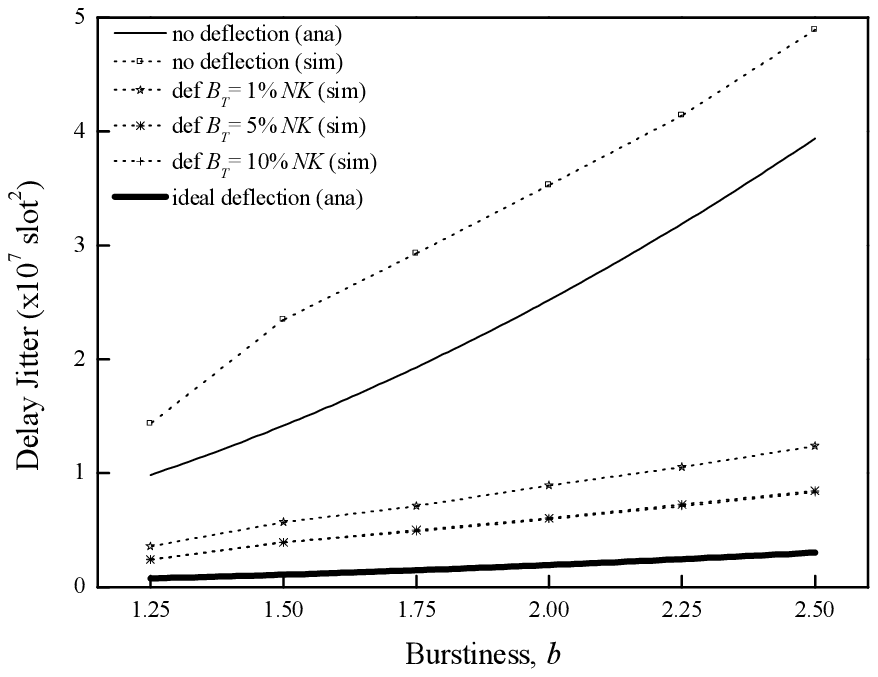}
\caption{Delay jitter versus burstiness, where $N=64$, $\hat{\lambda}=0.8$, $\bar{\lambda}=0.98/64$ and $K=\dot{K}$ or $\mathbf{\dot{K}}$.}\label{Jitter}
\end{figure}

The delay jitter versus burstiness is plotted in Fig. \ref{Jitter}, in which the delay variance $var[\mathbf{\dot{D}}]$ given by (\ref{EQ17}) provides a lower bound of the delay variance $var[\dot{D}]$ obtained from the discrete simulation model. Again, these curves clearly demonstrate that the delay jitter of the BvN switch without deflection is unsatisfactory when the burstiness $b$ of input traffic is high.

\section{Conclusion}\label{conclusion}
The quasi-static scheduling based on BvN decomposition is a virtual circuit (VC) switching technique for input-queued packet switches. Since the guaranteed capacity for each VC is estimated from the mean loading, as expected, the performance of BvN switches becomes unpredictable when the input traffic is bursty. On the other hand, deflection routing is a fundamental strategy for packet switched networks to eliminate heavy traffic and reduce the need for buffers. Combining these two techniques, the D-BvN switch architecture proposed in this paper provides a compromised scheduling solution for input-queued switches with VOQs. Our algorithm is specifically designed to adapt to the fluctuations of input traffic, and is very easy to implement in practice. This deflection-compensated scheme is devised to help offset the excessive buffer requirements caused by bursty traffic, and it can be easily extended to any network environment with capacity and QoS guarantees.

\appendices
\section{Fluid flow analysis of VC in D-BvN switches}\label{DBvN-ana}
The queuing behavior of a virtual circuit (VC) is analyzed in this appendix. As described in Section \ref{Model}, the arrival process is a Markov modulated on-off fluid flow, where the traffic rate is $\hat{\lambda}+\lambda_d$ during on periods and $\lambda_d$ during off periods, and the average traffic rate is $\bar{\lambda}+\lambda_d$. The input traffic state is denoted by $S$, the service rate of the VC is $C$, and the VOQ size is $K$.
In a VOQ, according to the standard fluid flow model illustrated in \cite{TTlee:Book2010}, the cumulative distribution function (CDF) of the queue length $x$ in on period $P_1(X)$ and in off period $P_0(X)$ satisfies the following equation:
\[\left\{
\begin{array}{l}
(\lambda_d-C)\times \displaystyle{\frac{dP_0(X)}{dX}}=-\beta P_0(X)+\alpha P_1(X)\\
(\hat{\lambda}+\lambda_d-C)\times \displaystyle{\frac{dP_1(X)}{dX}}=\beta P_0(X)-\alpha P_1(X)\\
P_1(0)=0\\
P_0(K)=\displaystyle{\frac{\alpha}{\alpha+\beta}}\\
\end{array}
\right..\]
It follows that the CDF of the queue length $x$ can be expressed by
\begin{equation}
\left\{
\begin{array}{l}
P_0(X)=A_0\alpha+A_1(\hat{\lambda}+\lambda_d-C)e^{-\varepsilon X}\\
P_1(X)=A_0\beta+A_1(C-\lambda_d)e^{-\varepsilon X}
\end{array}
(0\le X<K)\right.,\label{EQ23}
\end{equation}
where
\[A_0=\frac{-(C-\lambda_d)\alpha}{(\alpha+\beta)\left[-(C-\lambda_d)\alpha+(\hat{\lambda}+\lambda_d-C)\beta e^{-\varepsilon K}\right]},\]
\[A_1=\frac{\alpha\beta}{(\alpha+\beta)\left[-(C-\lambda_d)\alpha+(\hat{\lambda}+\lambda_d-C)\beta e^{-\varepsilon K}\right]},\]
and
\[\varepsilon=\frac{\alpha}{\hat{\lambda}+\lambda_d-C}-\frac{\beta}{C-\lambda_d}.\]
Differentiating (\ref{EQ23}), we obtain the following relevant probability density functions (PDFs):
\[\left\{
\begin{array}{l}
p_0(x=X)=-\varepsilon A_1(\hat{\lambda}+\lambda_d-C)e^{-\varepsilon X}\\
p_1(x=X)=-\varepsilon A_1(C-\lambda_d)e^{-\varepsilon X}
\end{array}
(0<X<K)\right.,\]
\[\textrm{Pr}\{x=0\}=A_0(\alpha+\beta)+A_1\hat{\lambda},\]
and
\begin{eqnarray*}
\textrm{Pr}\{x=K\}&=&1-\textrm{Pr}\{x\le K^-\}\\
       &=&1-[A_0(\alpha+\beta)+A_1\hat{\lambda}e^{-\varepsilon K}].
\end{eqnarray*}
Let $P_o$ be the overflow probability defined as follows:
\begin{eqnarray*}
P_o&=&\frac{\textrm{overflow traffic during period}\ T}{\textrm{input traffic during period}\ T}\\
&=&\frac{\textrm{Pr}\{X=K\}\times (\hat{\lambda}+\lambda_d-C)\times T}{(\bar{\lambda}+\lambda_d)\times T}.
\end{eqnarray*}
For a sufficiently long period of time $T$, the total amount of traffic entering the VOQ is $(\bar{\lambda}+\lambda_d)(1-P_o)T$, which can be classified into four classes as follows:\\
1) Traffic arriving at the VC when the VOQ is empty (i.e., $x=0$)

The VOQ is empty only when the traffic source is in the off state and the input traffic rate is $\lambda_d$. Therefore, the input traffic arriving at the VC when the VOQ is empty is $\lambda_d\times \textrm{Pr}\{x=0\}\times T$. Consequently, the probability that the waiting time $\mathbf{D}_q=0$ is
\begin{align*}
&\quad\ \textrm{Pr}\{\mathbf{D}_q=0\}\\
&=\frac{\textrm{traffic entering VC during}\ T\ \textrm{when}\ x=0}{\textrm{traffic entering VC over}\ T}\\
&=\frac{\lambda_d\times \textrm{Pr}\{x=0\}\times T}{(\bar{\lambda}+\lambda_d)(1-P_o)T}.
\end{align*}
2) Traffic arriving at the VC when the queue length of the VOQ is $x=X (0<X<K)$ and the input traffic state is on

In this case, the input traffic has to wait for $X/C$ before it can be served by the VC. Also, the input traffic rate is $\hat{\lambda}+\lambda_d$. Thus, the probability that the waiting time $\mathbf{D}_q=X/C$ and input traffic state is on is given by
\begin{flalign}
&\quad\ \textrm{Pr}\left\{\frac{X}{C}<\mathbf{D}_q<\frac{X+dX}{C},S=\textrm{on}\right\}&\nonumber\\
&=\small{\frac{\textrm{traffic entering VC during}\ T\ \textrm{when}\ 0<X<K\ \&\ S=\textrm{on}}{\textrm{traffic entering VC over}\ T}}&\nonumber\\
&=\frac{(\hat{\lambda}+\lambda_d)p_1(x=X)T}{(\bar{\lambda}+\lambda_d)(1-P_o)T}dX&\nonumber\\
&=\frac{(\hat{\lambda}+\lambda_d)p_1(x=\mathbf{D}_qC)T}{(\bar{\lambda}+\lambda_d)(1-P_o)T}Cd\mathbf{D}_q.&\nonumber
\end{flalign}
3) Traffic arriving at the VC when the queue length of the VOQ is $x=X (0<X<K)$ and the input traffic state is off

Similar with the previous case, the probability that the waiting time $\mathbf{D}_q=X/C$ and input traffic state is off is given by
\begin{flalign}
&\quad\ \textrm{Pr}\left\{\frac{X}{C}<\mathbf{D}_q<\frac{X+dX}{C},S=\textrm{off}\right\}&\nonumber\\
&=\small{\frac{\textrm{traffic entering VC during}\ T\ \textrm{when}\ 0<X<K\ \&\ S=\textrm{off}}{\textrm{traffic entering VC over}\ T}}&\nonumber\\
&=\frac{\lambda_d p_0(x=X)T}{(\bar{\lambda}+\lambda_d)(1-P_o)T}dX&\nonumber\\
&=\frac{\lambda_d p_0(x=\mathbf{D}_qC)T}{(\bar{\lambda}+\lambda_d)(1-P_o)T}Cd\mathbf{D}_q.&\nonumber
\end{flalign}
4) Traffic arriving at the VC when the queue length of the VOQ is $x=K$

The VOQ is full only when the traffic source is in the on state and the input traffic rate is $\hat{\lambda}+\lambda_d$. In this case, the rate of the traffic that actually enters the VOQ is $C$. Therefore, the probability that the waiting time of the input traffic is $K/C$ should be
\begin{align*}
&\quad\ \textrm{Pr}\left\{\mathbf{D}_q=\frac{K}{C}\right\}\\
&=\frac{\textrm{traffic entering VC during}\ T\ \textrm{when}\ x=K}{\textrm{traffic entering VC over}\ T}\\
&=\frac{C\times \textrm{Pr}\{x=K\}T}{(\bar{\lambda}+\lambda_d)(1-P_o)T}.
\end{align*}

From the probability distributions defined above, we can determine the first and second moments of the queuing delay that the traffic waits in the VOQ as follows:
\begin{flalign}
E[\mathbf{D}_q]&=0\times \frac{\lambda_d\times \textrm{Pr}\{x=0\} T}{(\bar{\lambda}+\lambda_d)(1-P_o)T}&\nonumber\\
&+\int_0^{\frac{K}{C}^-}\mathbf{D}_q\times&\nonumber\\
&\quad\frac{T [(\hat{\lambda}+\lambda_d)p_1(x=\mathbf{D}_qC)+\lambda_d p_0(x=\mathbf{D}_qC)]}{(\bar{\lambda}+\lambda_d)(1-P_o)T} Cd\mathbf{D}_q&\nonumber\\
&+\frac{K}{C}\times \frac{C\times \textrm{Pr}\{x=K\}T}{(\bar{\lambda}+\lambda_d)(1-P_o)T},&\label{EQ24}
\end{flalign}
and
\begin{flalign}
E[\mathbf{D}_q^2]&=0\times \frac{\lambda_d\times \textrm{Pr}\{x=0\} T}{(\bar{\lambda}+\lambda_d)(1-P_o)T}&\nonumber\\
&+\int_0^{\frac{K^-}{C}}\mathbf{D}_q^2\times&\nonumber\\
&\quad\frac{T [(\hat{\lambda}+\lambda_d)p_1(x=\mathbf{D}_qC)+\lambda_d p_0(x=\mathbf{D}_qC)]}{(\bar{\lambda}+\lambda_d)(1-P_o)T} Cd\mathbf{D}_q&\nonumber\\
&+\left(\frac{K}{C}\right)^2\times \frac{C\times \textrm{Pr}\{x=K\}T}{(\bar{\lambda}+\lambda_d)(1-P_o)T}.&\label{EQ25}
\end{flalign}

In the equilibrium state, when $N(\bar{\lambda}+\lambda_d)=NC$ and $K=\mathbf{\dot{K}}$, we obtain the following results:
\begin{flalign}
E[\mathbf{\dot{D}}_q]&=b\times &\nonumber\\ &\frac{\left[\left(2\frac{\alpha}{\beta}+1\right)\rho-\left(1+\frac{\alpha}{\beta}\right)\right]\left[\left(2-\frac{\beta}{\alpha}\right)\rho+\left(\frac{\beta}{\alpha}-1\right)\right]}{2\rho(1-\rho)},&\nonumber\\
\label{EQ26}
\end{flalign}
and
\begin{flalign}
E[\mathbf{\dot{D}}_q^2]=&b^2\times &\nonumber\\ &\frac{\left[\left(2\rho-1\right)\frac{\alpha}{\beta}-\left(1-\rho\right)\right]^2\left[\left(1-\rho\right)\frac{2\beta}{\alpha}+2\rho-1\right]}{3(1-\rho)^2},&\nonumber\\
\label{EQ27}
\end{flalign}

\section{Fluid flow analysis of VC in BvN switches}\label{BvN-ana}
The analytical results of the BvN switch without deflections are given in this appendix. We consider that the BvN switch under study is homogenous, such that all VCs are identical and independent to each other. The traffic input to each VC is a Markov modulated on-off fluid flow, where the traffic rate is $\hat{\lambda}$ in on periods and $0$ in off periods, and the average rate is $\bar{\lambda}$.

Similarly to Appendix \ref{DBvN-ana}, we firstly obtain the probability density functions (PDFs) of the queue length $x$ of each VOQ as follows:
\[\left\{
\begin{array}{l}
p_0(x=X)=-\varepsilon A_1(\hat{\lambda}-C)e^{-\varepsilon X}\\
p_1(x=X)=-\varepsilon A_1Ce^{-\varepsilon X}
\end{array}
(0<X<K)\right.,\]
\[\textrm{Pr}\{x=0\}=A_0(\alpha+\beta)+A_1\hat{\lambda},\]
and
\begin{equation}
\textrm{Pr}\{x=K\}=1-[A_0(\alpha+\beta)+A_1\hat{\lambda}e^{-\varepsilon K}].\label{EQ28}
\end{equation}
where
\[A_0=\frac{-C\alpha}{(\alpha+\beta)\left[-C\alpha+(\hat{\lambda}-C)\beta e^{-\varepsilon K}\right]},\]
\[A_1=\frac{\alpha\beta}{(\alpha+\beta)\left[-C\alpha+(\hat{\lambda}-C)\beta e^{-\varepsilon K}\right]},\]
and
\[\varepsilon=\frac{\alpha}{\hat{\lambda}-C}-\frac{\beta}{C}.\]

In BvN switches, overflow packets will be immediately dropped by VOQs. Following the same derivation procedure of $P_o$ and using (\ref{EQ28}), we can obtain the loss probability $P_l$ as follows
\[P_l=\frac{1}{\alpha+\beta}\frac{(\hat{\lambda}-C)\beta[C\alpha-\beta(\hat{\lambda}-C)]}{\bar{\lambda}[C\alpha e^{\varepsilon K}-\beta(\hat{\lambda}-C)]},\]
where
\[\varepsilon=\frac{\alpha}{\hat{\lambda}-C}-\frac{\beta}{C}.\]
On the other hand, if $P_l$ is given, the required VOQ buffer size can is given as follows
\begin{equation}
K=\frac{1}{\varepsilon}\ln\frac{\beta(\hat{\lambda}-C)+\displaystyle{\frac{(\hat{\lambda}-C)\beta[C\alpha-\beta(\hat{\lambda}-C)]}{(\alpha+\beta)\bar{\lambda}P_l}}}{C\alpha}.\label{EQ29}
\end{equation}

Using the same technique described in Appendix \ref{DBvN-ana}, we can obtain the average delay $E[D]$ and delay jitter $var[D]$ of BvN switches from (\ref{EQ29}) as follows:
\begin{flalign}
\qquad\qquad E[D]=&\int_0^{\frac{K}{C}^-}D\times \frac{\hat{\lambda}p_1(x=DC)}{\bar{\lambda}(1-P_l)}CdD&\nonumber\\
\qquad\qquad &+\frac{K}{C}\times \frac{C\times \textrm{Pr}\{x=K\}}{\bar{\lambda}(1-P_l)},&\nonumber\\
\label{EQ30}
\end{flalign}
and
\begin{flalign}
\qquad\qquad E[D^2]=&\int_0^{\frac{K}{C}^-}D^2\times \frac{\hat{\lambda}p_1(x=DC)}{\bar{\lambda}(1-P_l)}CdD&\nonumber\\
\qquad\qquad &+\left(\frac{K}{C}\right)^2\times \frac{C\times \textrm{Pr}\{x=K\}}{\bar{\lambda}(1-P_l)}.&\nonumber\\
\label{EQ31}
\end{flalign}

\section{Discrete Model of Simulation}\label{sim-model}
The time is slotted in the simulation model of D-BvN switches. We assume that the input stream to each VC is a discrete-time on-off process with geometrically distributed on and off states with transition probabilities $\alpha$ and $\beta$, respectively, and all VCs are statistically identical. Furthermore, we assume that an input packet is generated in each time slot with probability $\hat{\lambda}$ during the on state, $0<\hat{\lambda}\le1$, and no packets arrive during the off state. The probability $\hat{\lambda}$ is actually the peak arrival rate, such that the following average input rate
\[\bar{\lambda}=\frac{\hat{\lambda}\beta}{\alpha+\beta}<C=\frac{1}{N},\]
is consistent with the same parameter of the fluid-flow model employed in our analysis. As for BvN decomposition, we use a set of randomly generated $N$ circular-shift permutation matrices to guarantee that the average capacity assigned to each VC is $C=1/N$. In addition, we set the cross-switch constant delay a equal to one time slot in the simulation.

\bibliography{IEEEabrv,Bib_YT}
\end{document}